\documentclass[aps,twocolumn,pra,showpacs,amsmath,amssymb,showkeys,10pt]{revtex4-2}
\usepackage{graphicx}
\usepackage{epstopdf}
\usepackage[normalem]{ulem}
\usepackage{braket}
\usepackage{hyperref}
\usepackage[dvipsnames]{xcolor}
\usepackage{ulem}

\allowdisplaybreaks

\begin{document}

\title{Wave/particle duality in monitored Jaynes--Cummings resonances}

\author{Th. K. Mavrogordatos}
\email{themis.mavrogordatos@fysik.su.se}
 \affiliation{Department of Physics, AlbaNova University Center, SE 106 91, Stockholm, Sweden}

\date{\today}

\begin{abstract}
We operationally uncover aspects of wave/particle duality for the open driven Jaynes--Cummings (JC) model in its strong-coupling limit. We lay special emphasis on the vacuum Rabi resonance, and determine the corresponding normalized intensity-field correlation function via mapping to ordinary resonance fluorescence. We demonstrate that temporal wave-particle fluctuations of light emanating from an established vacuum Rabi resonance are explicitly non-classical, while the limit of vanishing spontaneous emission restores detailed balance. When photon blockade sets in, individual realizations show a rapidly increasing frequency of fluctuations towards the two-photon resonance, arising as a direct consequence of a resolved JC spectrum. About the two-photon resonance peak, spontaneous emissions are more likely to revive a high-frequency quantum beat in the conditioned electromagnetic field amplitude than photons escaping out of the cavity mode. The beat originates from a coherent superposition of the first excited JC couplet states, and sets the background against which nonclassical phase shifts are observed in the conditioned quadrature amplitudes. We also find that the onset of steady-state bimodality reduces the variation of the normalized intensity-field correlation, at the expense of its temporal symmetry.
\end{abstract}


\maketitle

\section{Introduction}

Steady states away from equilibrium are central to quantum optics. In the required open quantum system description, the correlation functions of light~\cite{Brown1956, Mandel1958, Glauber1963I, Glauber1963II, Rice2020} are measured via photoelectric detection on an outgoing field which is absorbed by the environment. In this context, fluctuations about a steady state, which is explicitly not symmetric under time reversal, may exhibit a specific time order different to the self-symmetric autocorrelation functions~\cite{Denisov2002}. The records obtained from quantum-trajectory theory disentangle the system and environment yet without discarding all correlations. At the same time, different environments give rise to mutually exclusive ways of record making, defining a self-consistent pure-state unraveling of the master equation~\cite{Carmichael1999}.

Conditional homodyne detection~\cite{GiantViolations2000,CarmichaelFosterChapter,Xu2015} is a distinctive method of record making which operationally reveals the wave/particle duality of quantum radiation. It cross-correlates the photocurrent of a phase-sensitive measurement with an initiating photon count, in what makes a natural extension of the technique introduced by Hanbury-Brown and Twiss~\cite{Brown1956, BrownHTwissI, BrownHTwissII} to determine the intensity (auto)correlation. The measured intensity-field correlations have been researched for the optical parametric oscillator (OPO)~\cite{GiantViolations2000}, as well as in both theoretical and experimental investigations of cavity quantum electrodynamics (cavity QED)~\cite{Foster2000, Reiner2001}. A temporal asymmetry in the third-order photon correlations has been experimentally reported for a two-photon resonance in the coherently driven Jaynes--Cummings (JC) model~\cite{Hamsen2017}, testifying to the breakdown of detailed balance~\cite{Koch2011}. Meanwhile, time-resolved oscillations at twice the light-matter coupling strength can be resolved when a laser excites the two-photon JC resonance, in the recent experiment by Najer and coworkers~\cite{Najer2019}. 

In this report, drawing principally on~\cite{WPCorr2024, JOSAB2024}, we employ the JC oscillator as a photoemissive source operating in the photon-blockade regime under strong-coupling conditions. We aim to contextually assess light-matter coupling through the cross-correlation of two different observables, photon number and field amplitude, and identify key differences in the system response arising between the vacuum Rabi and the two-photon resonances. Temporal asymmetry in the intensity-field correlation, computed as an average over trajectories generated via the wave-particle correlator unraveling, not only points to the breakdown of detailed balance, but also serves as a direct probe of non-Gaussian fluctuations~\cite{Denisov2002}. Operation at the vacuum Rabi resonance, on the other hand, will be shown to enjoy a special mapping to ordinary resonance fluorescence where quantum fluctuations are explicitly non-Gaussian but the intensity-field correlation is time symmetric~\citep{MarquinaCruz2008}. As soon as we move to different detunings this connection is lost, as exemplified by the asymmetric cross-correlations~\cite{Rice1988, Denisov2002} between cavity and spontaneous emissions at the two-photon resonance peak~\cite{JOSAB2024}. 

Salient features of the open coherently driven JC model in the photon blockade regime are briefly discussed in Sec.~\ref{sec:ME}, while the unraveling strategy is detailed in Sec.~\ref{sec:Wp}. The intensity-field correlation function is introduced in Sec.~\ref{sec:IF} as the main quantity of interest describing the averaged fluctuations of an electromagnetic wave amplitude conditioned on a photoelectron ``click'' in the steady state. Individual realizations--- extracted from the wave-particle correlator unraveling with variable drive-cavity detuning---associate the {\it conditioned} intensity-field correlations with the JC spectrum in the strong-coupling limit [Sec.~\ref{sec:CD}]. Concluding remarks close the paper out.

\begin{figure*}
\centering
\includegraphics[width=0.63\textwidth]{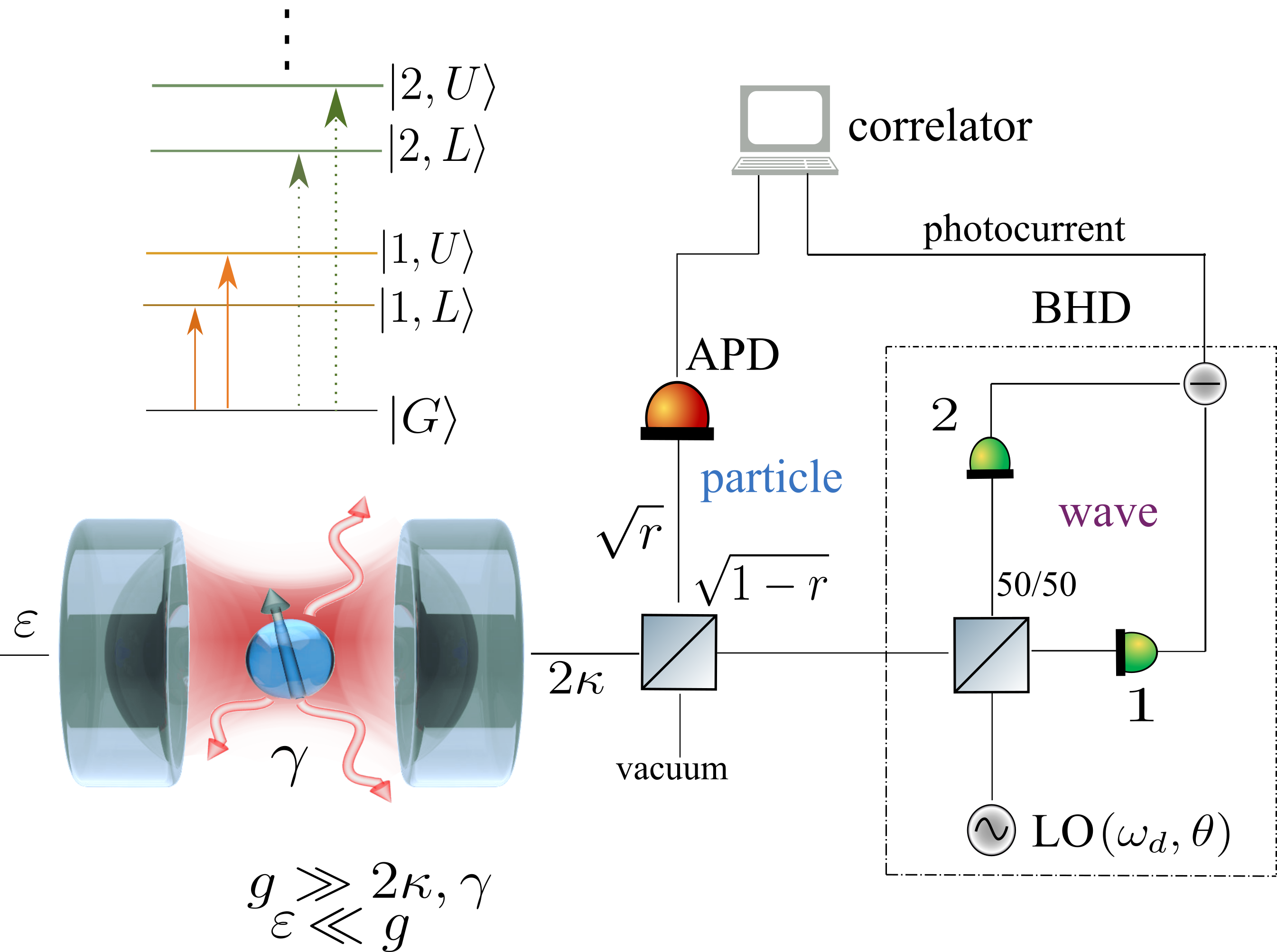}
\caption{ Schematic illustration of the wave-particle correlator admitting the Jaynes--Cummings (JC) oscillator in the photon blockade regime as a quantum source. A fraction $r$ of the input flux is sent to an avalanche photodiode (APD) in the ``start'' channel, while the remaining fraction $1-r$ is directed to a balanced homodyne detector (BHD) measuring the fluctuations of the signal field. The local oscillator is tuned to the drive frequency, while its phase $\theta$ selects the quadrature amplitude to be measured. Within a few correlation times following and preceding each ``start'' (an APD detection), the homodyne current $I_{\theta}(t)$ is recorded and added to a growing number of samples along an ongoing realization. Averaging over a large number of photocurrent samples reduces the shot noise so that the resulting signal is a conditional average of the quadrature amplitude fluctuations of the JC radiation input. The inset at the top left depicts the first four dressed JC states, comprising the ground ($|G\rangle$), the first-excited couplet ($|1,L\rangle$ and $|1,U\rangle$), and the the second-excited couplet ($|2,L\rangle$ and $|2,U\rangle$). The solid (dashed) arrows indicate a drive tuned to a vacuum Rabi (two-photon) resonance.}
\label{fig:FIG1}
\end{figure*} 

\section{Master equation and the Jaynes--Cummings photoemissive source}
\label{sec:ME}

For a single-`atom' cavity QED system, we employ the Lindblad master equation dictating the ensemble average evolution of the reduced system density matrix $\rho$~\cite{CarmichaelBook2}:
\begin{equation}\label{eq:ME}
\begin{aligned}
\frac{d\rho}{dt}=\mathcal{L}\rho=&\tfrac{1}{i\hbar}[H_{\rm JC},\rho]+\kappa (2a \rho a^{\dagger}-a^{\dagger}a \rho - \rho a^{\dagger}a)\\
&+(\gamma/2)(2\sigma_{-}\rho\sigma_{+}-\sigma_{+}\sigma_{-}\rho-\rho \sigma_{+}\sigma_{-}),
\end{aligned}
\end{equation}
where 
\begin{equation}\label{eq:JCHamiltonian}
H_{\rm JC}=-\hbar \Delta \omega(\sigma_{+}\sigma_{-} + a^{\dagger}a) + \hbar g (a\sigma_{+}+a^{\dagger}\sigma_{-}) + \hbar \varepsilon (a+a^{\dagger})
\end{equation}
is the Jaynes--Cummings (JC) Hamiltonian with coherent drive of amplitude $\varepsilon$, written in a frame rotating with the drive frequency $\omega_d$. In Eq.~\eqref{eq:JCHamiltonian}, $a^{\dagger}$ and $a$ are the raising and lowering operators for the intracavity mode of frequency $\omega_0$, $\sigma_{+}$ and $\sigma_{-}$ are the raising and lowering operators for the two-state `atom' on resonance with the field mode, and $g$ is the dipole coupling strength. In the ME~\eqref{eq:ME}, $2\kappa$ is the photon loss rate, and $\gamma$ is the spontaneous emission rate for the `atom'. The JC Hamiltonian models the quantum source in the strong-coupling limit, defined by $g\gg (2\kappa, \gamma)$ in line with recent developments in circuit QED~\cite{Blais2021}. We access the regime of multiphoton JC resonances with $\varepsilon/g \ll 1$ and a drive-cavity detuning $\Delta\omega\equiv (\omega_d-\omega_0) \sim g$. The field transmitted through the cavity mirrors forms the input to the wave-particle correlator, as shown in Fig.~\ref{fig:FIG1}. The eigenstates of the non-driven version of~\eqref{eq:JCHamiltonian} ($\varepsilon=\omega_d=0$) shape the familiar infinite ladder of excited state doublets~\cite{Carmichael2015}
\begin{equation}\label{eq:dressedstates}
\begin{aligned}
|n,L\rangle&=\frac{1}{\sqrt{2}}(|+\rangle |n-1\rangle - |-\rangle |n\rangle),\\
|n,U\rangle&=\frac{1}{\sqrt{2}}(|+\rangle |n-1\rangle + |-\rangle |n\rangle),
\end{aligned}
\end{equation}
alongside the ground state $|G\rangle=|-\rangle|0\rangle$, the common state out of which the two JC ladders begin. Here, $|+\rangle$ ($|-\rangle$) are the upper (lower) states of the two-state `atom', and $|n\rangle$ is the Fock state of the cavity field containing $n$ photons. 

The excited dressed states have energies
\begin{equation}
\begin{aligned}
E_{n,L}=n\hbar\omega-\sqrt{n}\hbar g,\\
E_{n,U}=n\hbar\omega+\sqrt{n}\hbar g,
\end{aligned}
\end{equation}
$n=1,2,\ldots$, revealing the characteristic $\sqrt{n}$ nonlinearity of the JC model~\cite{Fink2008, Bishop2009}. 

In the presence of the coherent drive but neglecting perturbative corrections (for $\varepsilon/g \ll 1$) owing to the so-called {\it dressing of the (bare) dressed states}, a multiphoton resonance of order $n$ is excited if the drive frequency satisfies the relation
\begin{equation}
n\hbar \omega_d=n\hbar \omega_0 \pm \sqrt{n}\hbar g,
\end{equation}  
in which case the detuning from the next ($n+1$) step up in the JC ladder of bare states~\eqref{eq:dressedstates} is~\citep{Carmichael2015}
\begin{equation}
E_{n+1,(U,L)}-E_{n,(U,L)}-\hbar\omega_d=\mp \left(\frac{n+1}{\sqrt{n}}-\sqrt{n+1}\right)\hbar g.
\end{equation}
When restricting the dynamics to a two-state manifold comprising $|G\rangle$ and $|1,U(L)\rangle$, a drive with detuning $\Delta\omega=\pm g$ excites a vacuum Rabi resonance. For a minimal four-state model comprising $|G\rangle$, $|1,L\rangle$, $|1,U\rangle$ and $|2,U(L)\rangle$, dressing of the bare dressed states modifies the drive detuning required for operation at the two-photon resonance to $\Delta\omega=\pm (g/\sqrt{2})[1+2(\varepsilon/g)^2]$--a perturbative second-order correction in the drive amplitude of the same order as the effective two-photon Rabi frequency~\cite{Shamailov2010, Lledo2021}.

\section{The wave-particle correlator unraveling of the master equation (1)}
\label{sec:Wp}

Let us now discuss a particular unraveling of the ME~\eqref{eq:ME}, which brings forward the dual aspect of the radiation emitted from a quantum source. The complementary unravelings~\cite{Carmichael1993QTII, Carmichael1999,Wiseman2012} are generated through the operation of the wave-particle correlator~\cite{GiantViolations2000,Foster2000,Carmichael2001,Reiner2001,CarmichaelFosterChapter} in the following fashion: photons (particles) trigger ``clicks'' in an avalanche photodiode (APD), which in turn reset conditioned records of electromagnetic field amplitudes (waves) in the photocurrent output of a balanced homodyne detector (BHD), as illustrated in Fig.~\ref{fig:FIG1}. Allowing for an unbalance $r$ (with $0\leq r \leq 1$) in the source photon flux directed to the two arms, the BHD receives as an input the portion of the field $2\sqrt{2\kappa (1-r)}\,A_{\theta}$, where 
\begin{equation}
A_{\theta} \equiv \tfrac{1}{2}(a e^{-i\theta}+a^{\dagger} e^{-i\theta})
\end{equation}
is the quadrature amplitude selected by the local oscillator (LO) phase $\theta$ (with $0\leq \theta <\pi$). For $r=1$, we revert to direct photodetection, the natural suggestion from the form of ME~\eqref{eq:ME}. For $r<1$, the charge $dq_{\theta}$ deposited on the detector circuit in the interval from $t$ to $t+dt$ generates the BHD photocurrent $I_{\theta}(t)$ satisfying $dI_{\theta}=-\tau_{d}^{-1}(I_{\theta}dt-dq_{\theta})$, with $\tau_d^{-1}$ the detection bandwidth. Using the JC oscillator as the quantum source, the detection bandwidth must on the order of (and bigger than) $g$ to accommodate the smallest timescale of the dynamics, which in turn requires a large number of photocurrent samples $N_s$ to recover the signal from the residual LO shot noise of variance $\propto 1/N_s$~\cite{GiantViolations2000, CarmichaelFosterChapter}. The field measured at the APD is proportional to $\sqrt{2\kappa r}\,a$, while detections occur with a probability equal to $p_{\rm APD}(t)=2\kappa r \langle \psi_{\rm REC}(t)| a^{\dagger}a|\psi_{\rm REC}(t)\rangle\,dt$, and the system wavefunction is updated to $\sqrt{2\kappa r} a|\psi_{\rm REC}\rangle$. Besides the ``start'' path---updating a cumulative average of photocurrent samples at every trigger of the APD---and the BHD, there is an additional output channel altogether: spontaneous emissions out of the sides of the cavity occur at the instantaneous rate $R_{\rm spon}(t)=\gamma\langle \psi_{\rm REC}(t)| \sigma_{+} \sigma_{-}|\psi_{\rm REC}(t)\rangle$, and update in their own right the system wavefunction to $\sqrt{\gamma}\sigma_{-}|\psi_{\rm REC}\rangle$. 

Between collapses, induced by spontaneous emission events and APD trigger ``clicks'', the un-normalized conditioned state $|\overline{\psi}_{\rm REC}\rangle$ satisfies the following stochastic Schr\"odinger equation (SSE)~\cite{Carmichael1993QTIII, Reiner2001, CarmichaelBook2}:
\begin{equation}\label{eq:SSEmain}
\begin{aligned}
d|\overline{\psi}_{\rm REC}\rangle=&\Big[\tfrac{1}{i\hbar}H_{\rm JC}\,dt-\kappa a^{\dagger}a\,dt-(\gamma/2)\sigma_{+}\sigma_{-}\,dt\\
&+\sqrt{2\kappa (1-r)}\, a\,e^{-i\theta} d\xi\Big]|\overline{\psi}_{\rm REC}\rangle,
\end{aligned}
\end{equation}
where
\begin{equation}\label{eq:dximain}
\begin{aligned}
d\xi&=\sqrt{2\kappa (1-r)}[(e^{i\theta}\langle a^{\dagger}  \rangle_{\rm REC} + e^{-i\theta} \langle a \rangle_{\rm REC}) dt] + dW\\
&=\sqrt{8\kappa (1-r)}\langle A_{\theta}\rangle_{\rm REC}\,dt + dW.
\end{aligned}
\end{equation}
In the above expression, $e$ is the electronic charge and $G$ is the detector gain; $dW$ is a Gaussian-distributed random variable with zero mean and variance $dt$. The two averages in Eq.~\eqref{eq:dximain} are to be calculated with respect to the normalized conditioned state 
\begin{equation}
|\psi_{\rm REC}(t)\rangle=\frac{|\overline{\psi}_{\rm REC}(t)\rangle}{\sqrt{\langle \overline{\psi}_{\rm REC}(t)|\overline{\psi}_{\rm REC}(t)\rangle}}.
\end{equation}
The conditioned wavefunction is propagated forwards in time via a numerically implemented Monte Carlo algorithm, detailed in Sec. 4B of~\cite{Reiner2001}.

\section{Intensity-field correlation and violation of classical bounds}
\label{sec:IF}

Having expanded on the unraveling procedure in Sec.~\ref{sec:Wp}, we proceed to define an ensemble averaged quantity, the normalized steady-state intensity-field correlation function~\cite{GiantViolations2000, Wiseman2002, CarmichaelFosterChapter,Xu2015}:
\begin{equation}\label{eq:hdef}
h_{\theta}(\tau) \equiv \frac{\langle : a^{\dagger}a(0) A_{\theta}(\tau) : \rangle_{\rm ss}}{\langle a^{\dagger}a \rangle_{\rm ss} \langle A_{\theta} \rangle_{\rm ss}}.
\end{equation}
This quantity operationally results from averaging the conditional homodyne photocurrent over many ``starts'' $N_s$ in order to diminish the residual shot noise~\cite{GiantViolations2000, CarmichaelFosterChapter}. Similar to the definition of the intensity-intensity correlation $g_{\rm ss}^{(2)}(\tau)$~\cite{CarmichaelBook1, CarmichaelBook2}, $h_{\theta}(\tau)$ is explicitly conditioned on a photoelectron ``click'', recored after the attainment of steady state, which sets the correlator of Fig.~\ref{fig:FIG1} in action~\cite{GiantViolations2000, Foster2000}.  

The numerator of Eq.~\eqref{eq:hdef} is calculated via the quantum regression formula~\cite{Denisov2002,CarmichaelFosterChapter,Xu2015} in the steady state:
\begin{equation}
2\langle : a^{\dagger}a(0) A_{\theta}(\tau) : \rangle_{\rm ss}=\begin{cases} {\rm tr}[a e^{\mathcal{L}\tau}(a\rho_{\rm ss} a^{\dagger})]e^{-i\theta} + {\rm c.c.}\quad \tau \geq 0\\
{\rm tr}[a^{\dagger}a e^{\mathcal{L}|\tau|}(a\rho_{\rm ss})]e^{-i\theta} + {\rm c.c.}\,\, \tau\leq 0,
 \end{cases}
\end{equation}
with the super-operator $\mathcal{L}$ defined in the ME~\eqref{eq:ME}, and the steady state solving $\mathcal{L} \rho_{\rm ss}=0$.

The dynamic Stark splitting is modelled by the two-state approximation~\cite{Tian1992, CarmichaelBook2}, in which the following non-Hermitian Hamiltonian governs the evolution between collapses:
\begin{equation}\label{eq:Heff}
H_{|\Delta\omega|=g}=\hbar (\varepsilon/\sqrt{2})(\ell_{+}+\ell_{-})-\tfrac{1}{2}i\hbar (\kappa+\gamma/2)\ell_{+}\ell_{-},
\end{equation}
where $\ell_{+}=|1,L\rangle \langle G|$ ($\ell_{+}=|1,U\rangle \langle G|$) for $\Delta\omega=g$ ($\Delta\omega=-g$), and $\ell_{-}=\ell_{+}^{\dagger}$. In this approximation, we map $a \to (1/\sqrt{2})\ell_{-}$ and $\sigma_{-} \to -(1/\sqrt{2})\ell_{-}$ when tuning to the lower Rabi resonance with $\Delta\omega=-g$, while $a \to (1/\sqrt{2})\ell_{-}$ and $\sigma_{-} \to (1/\sqrt{2})\ell_{-}$ when tuning to the upper Rabi resonance when $\Delta\omega=g$~\cite{Tian1992, CarmichaelBook2, Bishop2009}. The photon loss and spontaneous emission rates are now modified to $\kappa \langle \ell_{+}\ell_{-}(t)\rangle_{\rm REC}$ and $(\gamma/2) \langle \ell_{+}\ell_{-}(t)\rangle_{\rm REC}$, respectively, becoming equal for $\gamma=2\kappa$---an impedance matching condition between the two output channels. 

When we adopt a drive tuned to the lower or upper vacuum Rabi resonances, with $|\Delta\omega|=g$, the intensity-field correlation function pertaining to scattering into the cavity mode can be recast as a normal-ordered average in the form
\begin{equation}
h_{\theta}(\tau;|\Delta\omega|=g)=\frac{\langle : \ell_{+}\ell_{-}(0) [\ell_{-}(\tau)e^{-i\theta} + \ell_{+}(\tau)e^{i\theta}]:\rangle}{\langle \ell_{+}\ell_{-} \rangle_{\rm ss} \langle \ell_{-}e^{-i\theta} + \ell_{+}e^{i\theta} \rangle_{\rm ss}},
\end{equation}
where the averages are calculated in the two-state subspace $\{|G\rangle, |1,U(L)\rangle\}$ for $\Delta\omega=g$ ($\Delta\omega=-g$). The analytical expression for the intensity-field correlation in the two-state approximation~\cite{MarquinaCruz2008} reads:
\begin{equation}\label{eq:ifcorrTS}
\begin{aligned}
&h_{\pi/2}(|\tau|;|\Delta\omega|=g)\\
&=1-e^{-3 \overline{\gamma}|\tau|/4}\left[\cosh(\delta|\tau|) + \frac{\overline{\gamma}}{4\delta}(1-2Y^2)\sinh(\delta|\tau|)\right],
\end{aligned}
\end{equation}
in which $\overline{\gamma}=\gamma/2+\kappa$ is the effective decay rate from the dressed couplet state, $Y=2\varepsilon/\overline{\gamma}$ is the dimensionless drive parameter, and $\delta=(\overline{\gamma}/4)\sqrt{1-8Y^2}$ determines the effective Rabi frequency for strong driving when the effective two-state transition saturates. As dictated by the quantum regression formula~\cite{CarmichaelBook1}, the form of Eq.~\eqref{eq:ifcorrTS} is based on the evolution of the `atomic' polarization average $\langle \sigma_{-}(\tau)\rangle$ in resonance fluorescence~\cite{MarquinaCruz2008}, following ($\tau > 0$) or preceding ($\tau < 0$) a collapse to the lower state $|-\rangle$. Since the polarization is zero when the `atom' returns to its ground state, we expect $h_{\pi/2}(0;|\Delta\omega|=g)=0$. We also note that the intensity-field correlation is independent of the dipole coupling constant $g$, and exhibits the well known semiclassical oscillations (ringing) we meet in the intensity correlation function and the photoelectron waiting-time distribution of resonance fluorescence for a saturated transition~\cite{Carmichael1989, CarmichaelBook1}. In our example, the Rabi frequency $\sqrt{2}\varepsilon$ determining the ringing is two orders of magnitude lower than $g$.
\begin{figure*}
\centering
\includegraphics[width=0.89\textwidth]{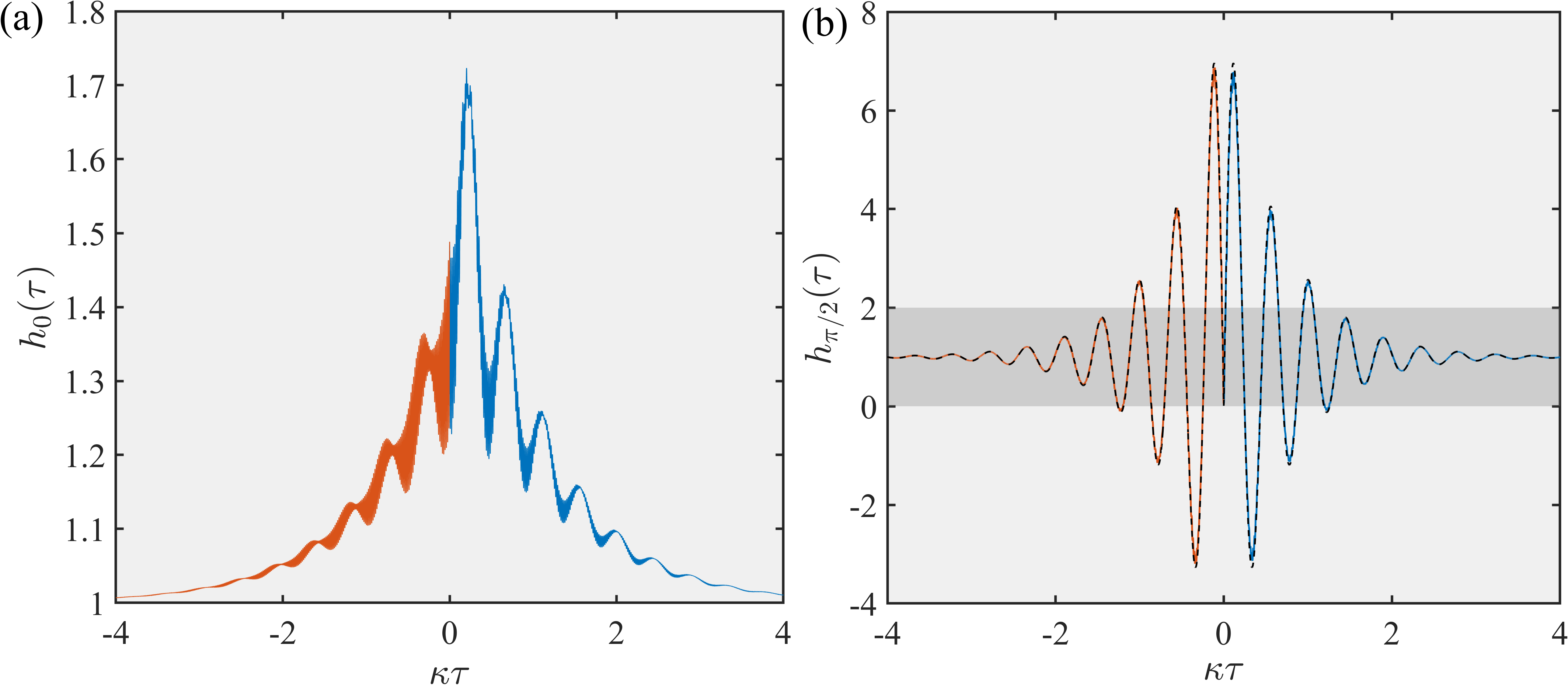}
\caption{Intensity-field correlation $h_{\theta}(\tau)$ against the dimensionless delay $\kappa \tau$, adopting a drive tuned to the upper vacuum Rabi resonance with $\Delta\omega=g$, numerically calculated via exact diagonalization of the Liouvillian along an exponential series expansion, and application of the quantum regression formula. Results are obtained using a Fock-state basis truncated at the twelve-photon level. In {\bf (a)}, we set $\theta=0$, and in {\bf (b)} we select $\theta=\pi/2$ in phase with the mean intracavity field. The dashed line in (b) plots the analytical expression~\eqref{eq:ifcorrTS}, while the shaded region indicates the values allowed for classical fields satisfying the inequality~\eqref{eq:B2}. The two different colours in (a, b) distinguish positive from negative time delays. The operating parameters substantiating the strong-coupling limit read: $g/\kappa=200$, $\varepsilon/g=0.05$ and $\gamma=2\kappa$.}
\label{fig:FIG2}
\end{figure*} 
\begin{figure*}
\centering
\includegraphics[width=0.88\textwidth]{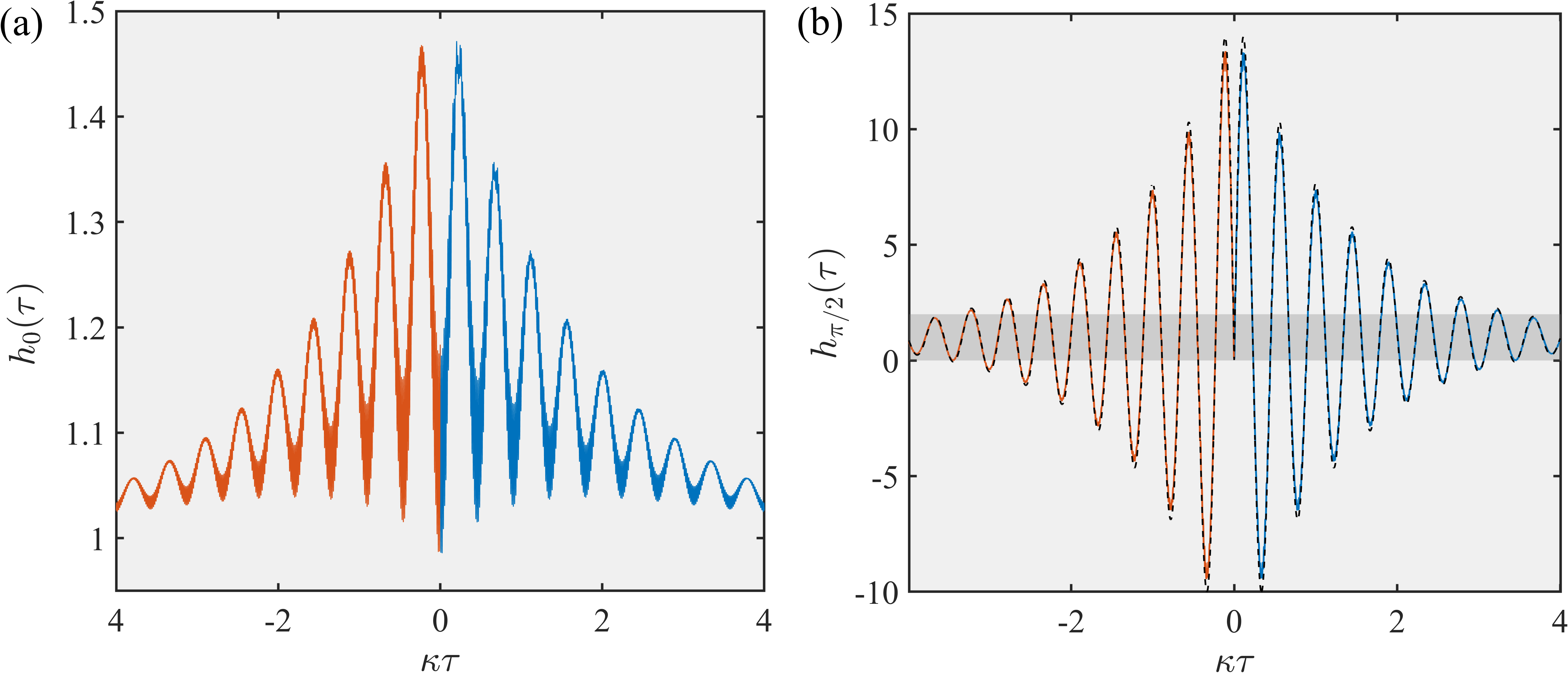}
\caption{Same as in Fig.~\ref{fig:FIG2}, but in the ``zero system size" limit $\gamma/(2\kappa) \to 0$.}
\label{fig:FIG3}
\end{figure*} 
\begin{figure*}
\centering
\includegraphics[width=0.99\textwidth]{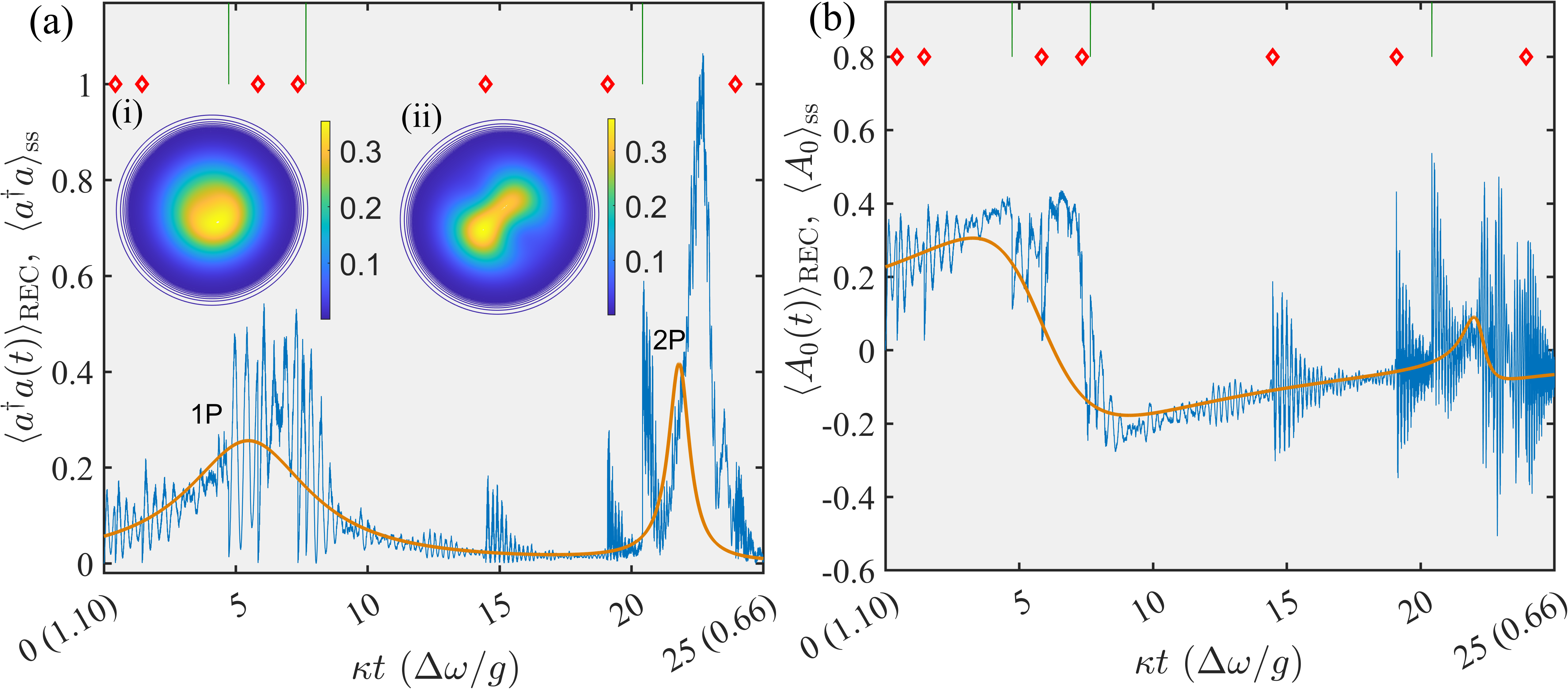}
\caption{Sample trajectory for a drive with variable detuning $\Delta\omega (t)$, traversing the vacuum Rabi and two-photon resonance peaks. The scaled detuning $\Delta\omega/g$ is uniformly scanned across the range $[1.10, 0.66]$ in the course of $25$ cavity lifetimes. {\bf (a)} Conditioned intracavity photon number $\langle \psi_{\rm REC}(t)|a^{\dagger}a|\psi_{\rm REC}(t)\rangle$ against the dimensionless time $\kappa t$, extracted from a Monte Carlo algorithm~\cite{Reiner2001} with collapses of two kinds (APD triggers and spontaneous emission events) and a continuous evolution in between obeying the SSE~\eqref{eq:SSEmain}. The LO phase is set to $\theta=0$. Diamonds mark spontaneous emissions out of the sides of the cavity, and top vertical strokes indicate the APD detection of a photon escaping out of the cavity mode. The thick solid line plots the photon number $\langle a^{\dagger}a \rangle_{\rm ss}$ determined from the numerically extracted steady-state solution of the ME~\eqref{eq:ME}. The left and right insets depict schematic contour plots of the cavity state Wigner distribution $W_{\rm ss}(x+iy)$ (with $-2\leq x,y \leq 2$) at the peak of the vacuum Rabi resonance (1P) and the two-photon resonance (2P), respectively. {\bf (b)} Conditioned quadrature amplitude $\langle \psi_{\rm REC}(t)|A_{\theta}|\psi_{\rm REC}(t)\rangle$, against the dimensionless time $\kappa t$, in the same realization as in frame (a) extracted from the aforementioned Monte Carlo procedure with the LO phase set to $\theta=0$. The thick solid line depicts $\langle A_{\theta} \rangle_{\rm ss}={\rm Re}[\langle a\rangle_{\rm ss} e^{-i\theta}]$ for $\theta=0$, while $\langle a\rangle_{\rm ss}$ is computed from the numerical steady-state solution of the ME~\eqref{eq:ME}. Operating parameters in (a, b) read: $g/\kappa=200$, $\varepsilon/g=0.05$ and $\gamma=2\kappa$. The wave-particle correlator functions with $r=0.5$. The initial state is $|G\rangle$, and the Fock-state basis is truncated at the twelve-photon level. }
\label{fig:FIG4}
\end{figure*} 
\begin{figure*}
\centering
\includegraphics[width=0.99\textwidth]{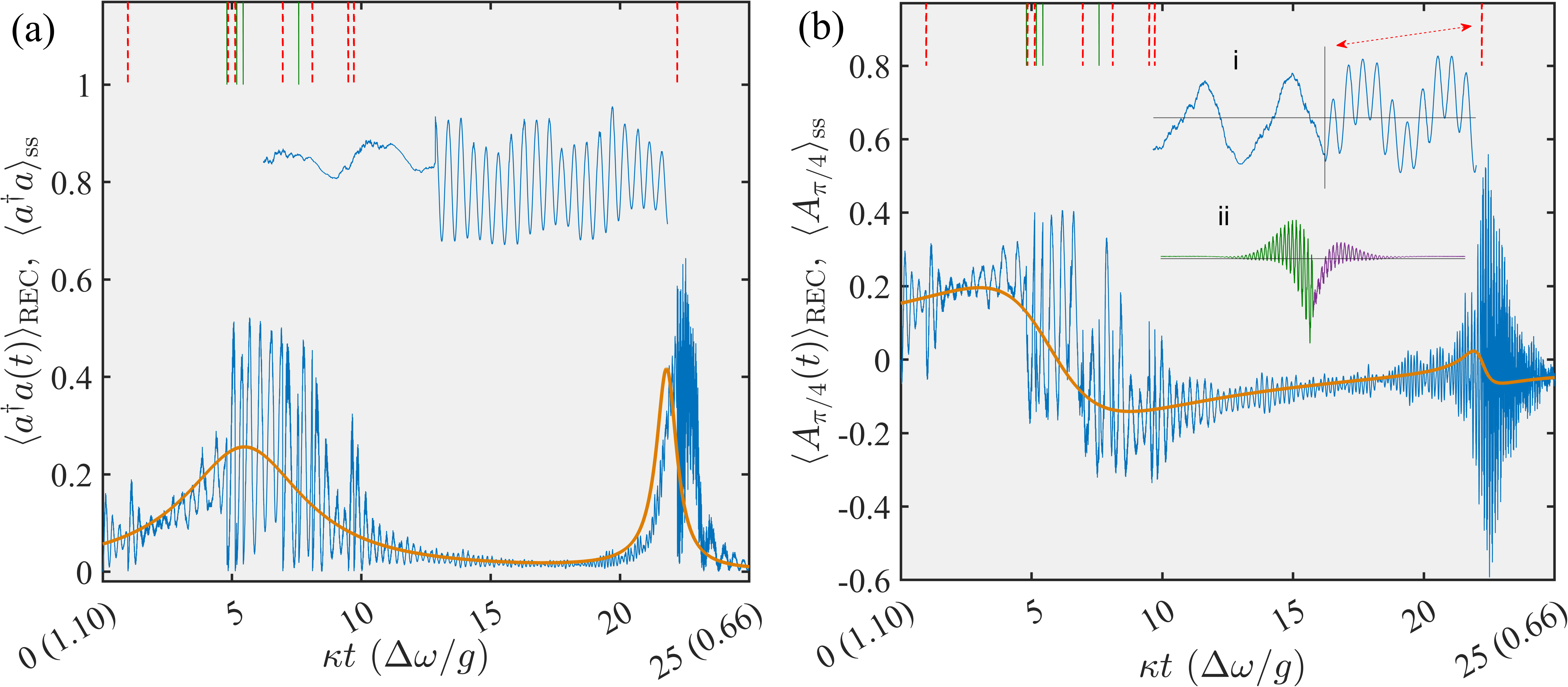}
\caption{Same as in Fig.~\ref{fig:FIG4}, but for the LO phase set to $\pi/4$. Dashed (in red) and solid (in green) strokes on top indicate spontaneous emissions and APD triggers, respectively. The inset in (a) focuses on the revival of the quantum beat in a window of $0.5$ cavity lifetimes about the last spontaneous emission. Inset {\bf i} in (b) focuses on a time window of $0.4$ cavity lifetimes centred about the last spontaneous emission event at $\kappa t_{\rm spon}\approx 22.21$ indicated by the thin vertical line, while inset {\bf ii} depicts $h_{\pi/4}(\tau)$ across 4 photon lifetimes for the detuning $\Delta\omega/g=0.7096$ corresponding to $\kappa t_{\rm spon}$, when the last spontaneous emission is recorded. The horizontal line indicates the zero level.}
\label{fig:FIG5}
\end{figure*} 
\begin{figure*}
\centering
\includegraphics[width=0.99\textwidth]{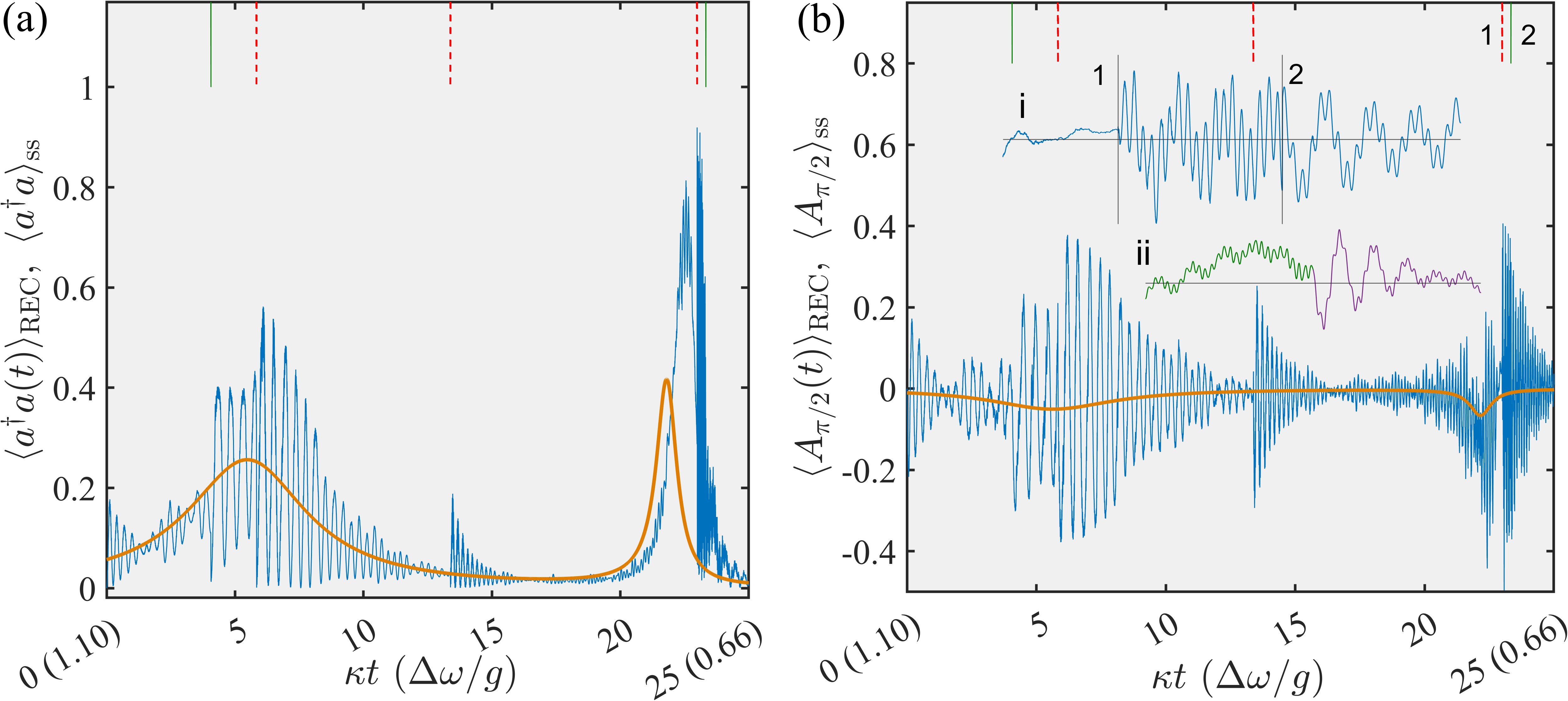}
\caption{Same as in Fig.~\ref{fig:FIG5}, but for $\theta=\pi/2$. Inset {\bf i} in (b) focuses on a time window of about $0.9$ cavity lifetimes including the last pair of emissions, at $\kappa t_{\rm spon}\approx 22.99$ and $\kappa t_{\rm APD}\approx 23.33$, indicated by the vertical lines 1 and 2, respectively. Inset {\bf ii} plots $h_{\pi/2}(\tau)$  across half a photon lifetime for the detuning $\Delta\omega/g=0.6900$ corresponding to $\kappa t_{\rm APD}$, when the last cavity emission is recorded by the APD. The horizontal lines mark the zero level.}
\label{fig:FIG6}
\end{figure*} 
\begin{figure*}
\centering
\includegraphics[width=0.99\textwidth]{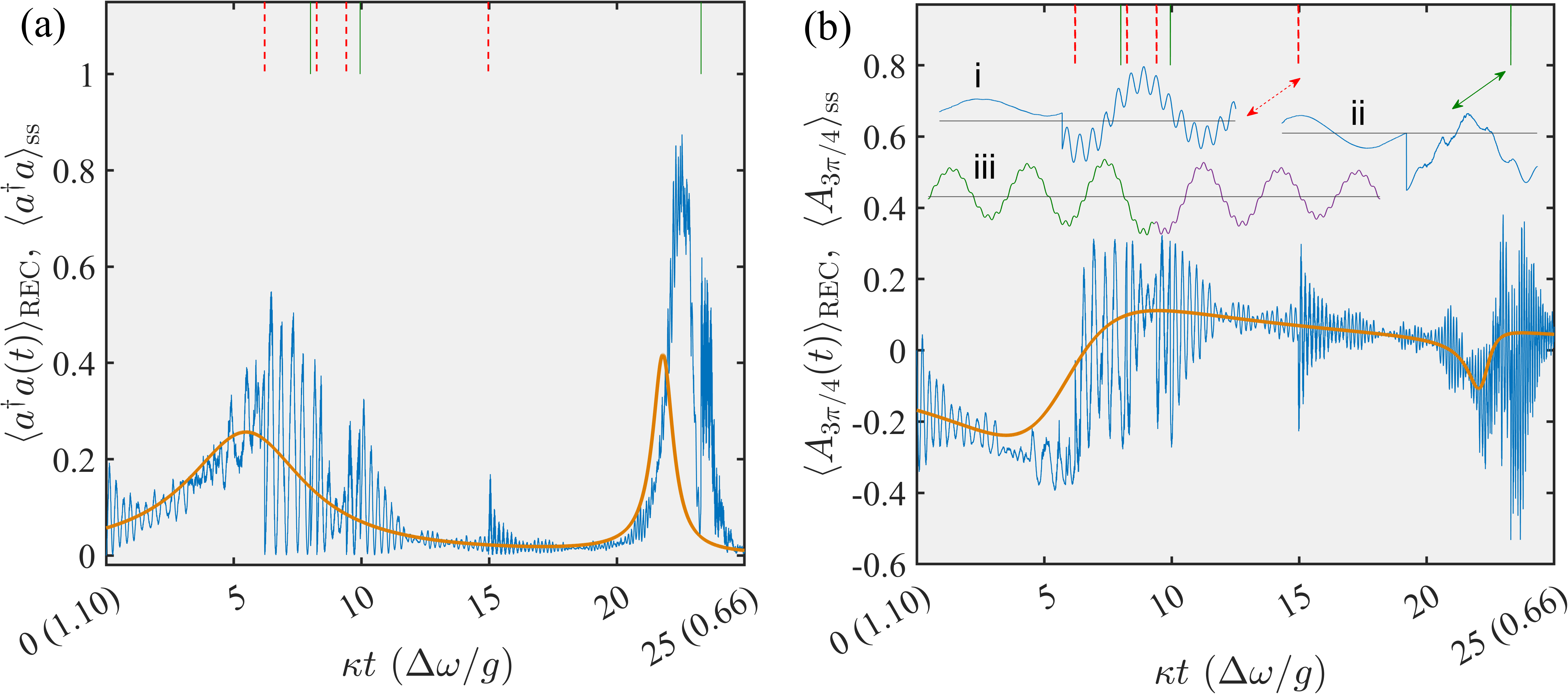}
\caption{Same as in Fig.~\ref{fig:FIG5}, but for $\theta=3\pi/4$. Inset {\bf i} in (b) focuses on a time window of $0.4$ cavity lifetimes at $\kappa t_{\rm spon}\approx 14.97$, when a spontaneous emission is recorded, while inset {\bf ii} plots $h_{3\pi/4}(\tau)$ across half a photon lifetime for the detuning $\Delta\omega/g=0.8370$ corresponding to $\kappa t_{\rm spon}$. Inset {\bf iii} focuses on a window of $0.2$ cavity lifetimes about $\kappa t_{\rm APD}\approx 23.30$, when the last cavity emission is recorded by the APD. The two emission events are indicated by the corresponding double-headed arrows. Horizontal lines mark the zero level.}
\label{fig:FIG7}
\end{figure*} 

\begin{figure*}
\centering
\includegraphics[width=0.99\textwidth]{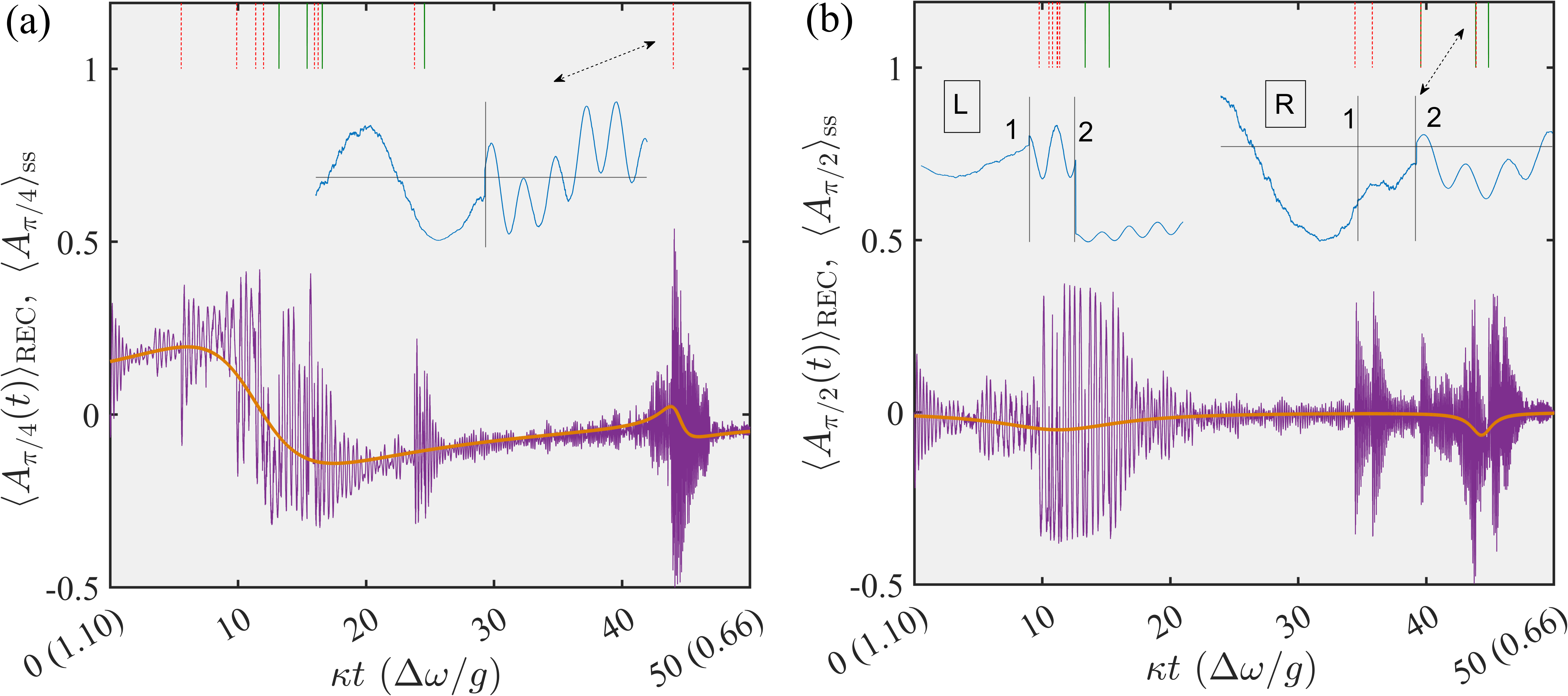}
\caption{Sample trajectories for a drive with variable detuning $\Delta\omega (t)$, traversing the vacuum Rabi and two-photon resonance peaks. The scaled detuning $\Delta\omega/g$ is uniformly scanned across the range $[1.10, 0.66]$ in the course of $50$ cavity lifetimes. Conditioned quadrature amplitude $\langle \psi_{\rm REC}(t)|A_{\theta}|\psi_{\rm REC}(t)\rangle$ obtained upon setting the LO phase to {\bf (a)} $\theta=\pi/4$, and {\bf (b)} $\theta=\pi/2$, against the dimensionless time $\kappa t$. The amplitudes are obtained in the course of two different realizations of the Monte Carlo procedure, one for each value of $\theta$. The thick solid line depicts $\langle A_{\theta} \rangle_{\rm ss}={\rm Re}[\langle a\rangle_{\rm ss} e^{-i\theta}]$ for $\theta=\pi/4$ and $\pi/2$ in (a), (b), respectively, while $\langle a\rangle_{\rm ss}$ is computed from the numerical steady-state solution to the ME~\eqref{eq:ME}. The inset in (a) focuses on a time window of $0.2$ cavity lifetimes about the last spontaneous emission. Inset L in frame (b) depicts the conditional photon number $\langle \psi_{\rm REC}(t)|a^{\dagger}a|\psi_{\rm REC}(t)\rangle$, focusing on the time window of $[43.80, 43.95]$ (in cavity lifetimes) about the closely-spaced pair of emissions indicated by the double-headed arrow. The cavity (spontaneous) emission event is marked by 1 (2). Inset R depicts the conditional amplitude $\langle \psi_{\rm REC}(t)|A_{\pi/2}|\psi_{\rm REC}(t)\rangle$ in the same time window as in inset L. The two vertical lines indicate the emission times, while the horizontal lines indicate zero amplitude. Operating parameters in both frames read: $g/\kappa=200$, $\varepsilon/g=0.05$ and $\gamma=2\kappa$. The wave-particle correlator functions with $r=0.5$. The initial state is the JC ground state, $|G\rangle$, and the Fock-state basis is truncated at the twelve-photon level.}
\label{fig:FIG8}
\end{figure*} 

In the two-state approximation with Hamiltonian~\eqref{eq:Heff}, the steady-state average of the signal mode is purely imaginary, as is $\langle \sigma_{-} \rangle_{\rm ss}$ in resonance fluorescence with system Hamiltonian $H^{\prime}=\hbar (\varepsilon/\sqrt{2})(\sigma_{+}+\sigma_{-})$. Hence, adjusting the LO in phase with the steady-state field by selecting $\theta=\pi/2$, the classical bounds for $h_{\pi/2}(|\tau|;|\Delta\omega|=g)$ are set by the inequalities~\cite{GiantViolations2000, CarmichaelFosterChapter, MarquinaCruz2008} 
\begin{subequations}
\begin{align}
&0 \leq h_{\pi/2}(0;|\Delta\omega|=g)-1\leq 1, \label{eq:B1}\\
|h_{\pi/2}(|\tau|&;|\Delta\omega|=g)-1|\leq |h_{\pi/2}(0;|\Delta\omega|=g)-1| \leq 1. \label{eq:B2}
\end{align}
\end{subequations} 
We note that $h_{\pi/2}(0;|\Delta\omega|=g)=0$, violating~\eqref{eq:B1} for every value of the drive amplitude. Inequality~\eqref{eq:B2} is also violated whenever the correlation attains negative vales, as is the case in Figs.~\ref{fig:FIG2}(b) and~\ref{fig:FIG3}(b). The two-state approximation predicts $h_{0}(|\tau|;|\Delta\omega|=g)=0$, since the average polarization $\langle \sigma_{-}(\tau)\rangle$ evolving from the lower state remains purely imaginary throughout. The numerical solution~\cite{Tan1999} of the ME~\eqref{eq:ME} in the steady state combined with the quantum regression formula~\footnote{Simulations are performed in \textsc{Matlab}'s {\it Quantum Optics Toolbox} using exact diagonalization along an exponential series expansion of the Liouvillian $\mathcal{L}$ in Eq.~\eqref{eq:ME}.}, however, produces a non-zero correlation $h_{0}(|\tau|;|\Delta\omega|=g)$, as we can see in Figs.~\ref{fig:FIG2}(a) and~\ref{fig:FIG3}(a), with the same number of semi-classical oscillations completed in four average photon lifetimes as for $\theta=\pi/2$. Excitation of the JC oscillator to the second-excited couplet states ($|2,U(L)\rangle$) brings in a quantum beat of frequency $\approx 2g$~\cite{Tian1992, Shamailov2010, Lledo2021}, and breaks detailed balance. Further along, in Sec.~\ref{sec:CD}, we meet with conditional amplifications of the quantum beat, instigated by tractable emission events. 

The temporal symmetry of $h_{0}(|\tau|;|\Delta\omega|=g)$ is restored upon approaching $\gamma/(2\kappa)\to 0$, the so-called limit of ``zero system size'' conserving the length of the Bloch vector~\cite{Alsing1991, CarmichaelBook2, Carmichael2015}. We remark that this limit is of special significance to the open driven JC model in the photon blockade regime, owing to the continual disagreement between the quantum fluctuations and the mean-field nonlinearity. The disparity is noted as an emerging system size parameter $[g/(2\kappa)]^2$, different to that of absorptive optical bistability~\cite{Savage1988}, is sent to infinity~\cite{Carmichael2015, SC2019}. Numerical simulations show that, for $\gamma/(2\kappa) \to 0$, intensity-field correlations are time-symmetric for every value of the LO phase $\theta$. Moreover, the numerically computed correlation function $h_{\pi/2}(|\tau|;|\Delta\omega|=g)$ is in very good agreement with the analytical expression of Eq.~\eqref{eq:ifcorrTS} for all ratios $\gamma/(2\kappa)$ up to the ``zero system size'', as exemplified in Figs.~\ref{fig:FIG2}(b) and~\ref{fig:FIG3}(b).

\section{Complementary conditioned vs. ensemble average dynamics}
\label{sec:CD}

To meet our goal, as set out in the Introduction, we must turn to the {\it conditioned} dynamical evolution dictated by an unraveling accomplished by the wave-particle correlator, when equal amounts of the incoming photon flux are directed to its two arms. We do so by adding a new element to the picture: a system parameter is varied to demonstrate that the addition or subtraction of just one quantum, or maybe a few, is noticeable in the obtained records as observable physics. It arises as a consequence of the strong-coupling limit, uncovering a well defined JC spectrum against the widths set by dissipation~\cite{CarmichaelBook2}. In Figs.~\ref{fig:FIG4}--\ref{fig:FIG8}, the conditional photon-number and quadrature amplitude expectations are plotted against $\Delta\omega(t)/g$ while the drive detuning is scanned ``fast'' (with a scan rate of $\approx 3.5g$ per $g^{-1}$) with the drive amplitude remaining constant and for impedance matching conditions in the output channels. In the course of $25$ cavity lifetimes ($50$ in Fig.~\ref{fig:FIG8}), the vacuum Rabi and two-photon resonance peaks are traced out for four complementary unravelings of the ME~\eqref{eq:ME}, corresponding to four different settings of the LO phase. The values of $\theta$ selected are characteristic to the nonlinearity displayed on the Wigner function~\cite{CarmichaelBook1} profiles of the cavity state, $W_{\rm ss}(x+iy)$. On top of each quantum trajectory are marked the times of spontaneous emission events along the photodetections at the APD in the ``start'' channel; the latter reset the sample making of $I_{\theta}(t)$. Moreover, ensemble averages obtained from the steady-state solution of the ME~\eqref{eq:ME} are overlaid upon the sample trajectories that unravel it, for the corresponding values of the cavity detuning. For the selected value of the drive amplitude, all steady-state averages evince a saturated vacuum Rabi resonance with a peak $\langle a^{\dagger}a \rangle_{\rm ss}\approx 1/4$ at $\Delta\omega=g$, as well as a two-photon resonance peak located at $\Delta \omega\approx (g/\sqrt{2})[1+2(\varepsilon/g)^2]$. 

The sample realization generated when setting $\theta=0$ corresponds to the intensity-field correlation function we have already met in Fig.~\ref{fig:FIG2}(a); in the present instance, however, larger manifolds of dressed states along the JC ladder progressively participate in the {\it conditioned} dynamics. In Fig.~\ref{fig:FIG4}(a), the first thing to observe is that the regression of fluctuations from cavity and spontaneous emissions takes place at an increasing frequency as the detuning traverses the region between the two resonance peaks. The semiclassical ringing at the frequency $\approx \sqrt{2}\varepsilon$ gives its place to a quantum beat of frequency $\approx 2g$, evincing a conditional coherent superposition of the states $|1, L\rangle$ and $|1,U\rangle$~\cite{WPCorr2024}. Another feature of note is that the conditioned intracavity photon number, and along with it the photon scattering rate $R_{\rm APD}(t)=2\kappa r \langle \psi_{\rm REC}(t)| a^{\dagger}a|\psi_{\rm REC}(t)\rangle$ in the ``start'' channel, peaks at a different detuning with more than double the amplitude of the steady-state maximum value. Simultaneously, as the detuning nears the two-photon resonance region, another longer timescale manifests. Right at the two-photon resonance peak, the two-photon Rabi frequency is perturbatively calculated to $2\sqrt{2}\varepsilon^2/g$~\cite{Najer2019, Lledo2021}. This semiclassical timescale, the longest in the displayed dynamical evolution, is the one to largely dictate the regression of fluctuations following the last pair of emissions, a cavity and a spontaneous emission either side of the resonance peak. The conditioned field amplitude plotted in Fig.~\ref{fig:FIG4}(b) also displays fluctuations of increasing frequency and co-existing timescales, while averaging in sign to the variations of the steady-state value of the quadrature amplitude. 

Let us further explore the correlations between photon ``clicks'' and the field amplitudes they condition, and vice versa. Next, we align the LO phase $\theta$ to the direction of developing bimodality in the phase portrait of the two-photon resonance peak, the small-amplitude nonlinearity evinced by the Wigner function in the right inset of Fig.~\ref{fig:FIG4}(a). Individual records about the peak systematically reveal that an emission is required in order to match the conditioned instantaneous photon scattering rate $R_{\rm APD}(t)=2\kappa r \langle \psi_{\rm REC}(t)| a^{\dagger}a|\psi_{\rm REC}(t)\rangle$ with the steady-state response. For the sample trajectory shown in Fig.~\ref{fig:FIG5}(a), the last spontaneous emission instigates intense oscillations at the quantum beat superimposed on top of the semiclassical ringing. Upon the revival of the quantum beat, the conditioned quadrature phase amplitude maintains its sign, while the amplitude $\langle A_{\pi/4}\rangle_{\rm ss}$ depicted in Fig.~\ref{fig:FIG5}(b) attains a negative value of small magnitude at the corresponding detuning. This observation is consistent with the large negative value of the normalized intensity-field correlation at zero delay. The quantum beat dephases over different realizations at a particular detuning, making only a small superposition to slower oscillations in the time-asymmetric $h_{\pi/4}(\tau)$. The frequency of these oscillations is related to the disparity between the effective dissipation rates in the two paths mediating the two-photon decay; they come about as a deviation from the secular approximation~\cite{Shamailov2010, WPCorr2024}.

\begin{figure*}
\centering
\includegraphics[width=0.99\textwidth]{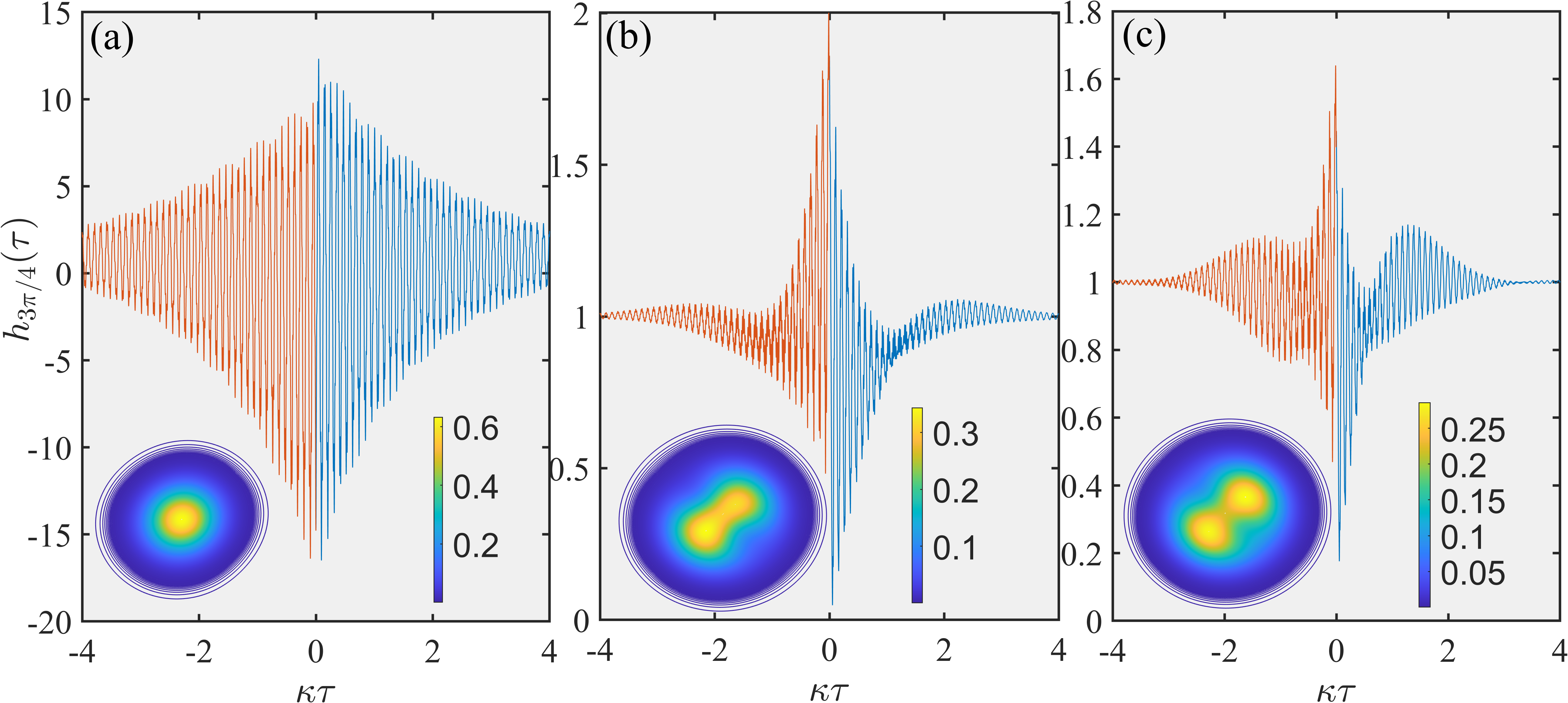}
\caption{Intensity-field correlation function $h_{\theta}(\tau)$ against the dimensionless delay $\kappa \tau$, numerically calculated from the steady-state solution of ME~\eqref{eq:ME} and the quantum regression formula, adopting a drive exciting the upper two-photon resonance with detuning $\Delta\omega=(g/\sqrt{2})[1+2(\varepsilon/g)^2]$. The LO phase is chosen perpendicular to the direction of steady-state bimodality, at $\theta=3\pi/4$. The scaled drive amplitude $\varepsilon/g$ is $0.01$ in {\bf (a)}, $0.04$ in {\bf (b)} and $0.05$ in {\bf (c)}. The three insets depict schematic contours of the corresponding steady-state Wigner functions $W_{\rm ss}(x+iy)$ (with $-2\leq x,y \leq 2$). A Fock-state basis is truncated at the twelve-photon level. The remaining operating parameters read: $g/\kappa=200$ and $\gamma/(2\kappa) \to 0$.}
\label{fig:FIG9}
\end{figure*} 

The complementary unraveling attained for $\theta=\pi/2$ and illustrated in Fig.~\ref{fig:FIG6} is produced with the LO in phase with the steady-state amplitude at the vacuum Rabi resonance. As such, it corresponds to the intensity-field correlations of Figs.~\ref{fig:FIG2}(b) and~\ref{fig:FIG3}(b). The penultimate emission out of the sides of the cavity instigates an intense revival of the quantum beat, which in turn periodically increases the conditioned probability for another emission, with frequency $\sim g$. Indeed, as we can see in Fig.~\ref{fig:FIG6}(a), a cavity photon is recorded by the APD within a fraction of the average photon lifetime, substantiating photon bunching in the operational sense~\citep{Carmichael1993QTII}. The steady-state electromagnetic field amplitude remains very close to zero [Fig.~\ref{fig:FIG6}(b)], yet the fluctuating conditioned amplitude exhibits the familiar chirp pattern when the second excited JC couplet is progressively accessed. The fact that $h_{\pi/2}(\tau=0)=0$ at the vacuum Rabi resonance implies that over an ensemble of realizations there are as many positive as there are negative field amplitude values conditioned on emission events, in those occasions that make a departure from the two-state model [see also the trajectory of Fig. 1 (lower frame) in~\cite{Tian1992}, obtained for direct photodetection ($r=1$)]. Towards the end of the trajectory, we meet with an instance where a positive quadrature amplitude is conditioned on the last two emissions, while the steady-state amplitude is negative and $h_{\pi/2}(0)>0$. To interpret the apparent contradiction in this particular sample realization of bunching, we must also take into account the fast variation of $h_{\pi/2}(\tau)$ about zero, at the detuning considered; the correlation function changes sign within a quantum beat of frequency $\sim g$, a characteristic variation time which cannot be resolved by the ``fast'' scan of the detuning. 

The last in the series of complementary unravelings has a realization generated for $\theta=3\pi/4$, the direction perpendicular to the onset of bimodality at a two-photon resonance. In Fig.~\ref{fig:FIG7}(a), we note that a spontaneous emission instigates large-amplitude oscillations of frequency $\approx \sqrt{2}\varepsilon$ at the peak of the vacuum Rabi resonance, while on the other hand the last cavity emission of the trajectory is followed by fluctuation regression about the steady-state two-photon resonance peak, albeit at a different detuning. In Fig.~\ref{fig:FIG7}(b), the $\pi$ phase change of the conditioned field amplitude at the last spontaneous emission is in agreement with a negative zero-delay intensity-field correlation, where there is no sign change until several quantum beats have lapsed. The same is true for the last cavity emission and the corresponding $h_{3\pi/4}(\tau)$. 

We might wonder at this point how the conditional wave-particle correlations pan out at a ``slower'' scan of the drive-cavity detuning (halving the scan rate to $\approx 1.75g$ per $g^{-1}$). Figures~\ref{fig:FIG8}(a) and~\ref{fig:FIG8}(b) are to be compared against Figs.~\ref{fig:FIG5}(b) and~\ref{fig:FIG6}(b), respectively, depicting the fluctuating quadrature amplitude along two characteristic directions (LO phases) in the steady-state phase space profile at the vacuum Rabi and the two-photon resonance peaks [see the Wigner functions of Fig.~\ref{fig:FIG4}(a)]. For $\theta=\pi/4$ [Fig.~\ref{fig:FIG8}(a)], no visible change is noticed. All but one emission events occur about the vacuum Rabi resonance, while the last photoelectron ``click'' due to a photon emitted out of the sides of the cavity instigates a discontinuity of the field towards positive amplitudes---a consequence of $h_{\rm \pi/4}(0)>0$---while reviving the quantum beat. Notable changes arise, however, when $\theta=\pi/2$ in the trajectory of Fig.~\ref{fig:FIG8}(b). First, the emissions events treble with respect to the trajectory of Fig.~\ref{fig:FIG6}(b), being evenly spread among the two resonance peaks. A more subtle difference also emerges, concerning the operational manifestation of photon bunching in a detuning regime where the zero-delay intensity correlation function is above unity [$g_{\rm ss}^{(2)}(0)>1$]: there is now a sign change in the conditioned field amplitude following two closely spaced emissions. The jumps are recoded much closer together when the detuning scan rate is halved, measured against the quantum beat which serves as a fluctuation `clock'. The two insets of Fig.~\ref{fig:FIG8}(b) show that the two emissions about $\kappa t=43.87$ are less than two beat cycles apart. While the beat emerges in the conditioned photon excitation immediately after the APD ``click'' (first emission from the pair), it does so for the field amplitude only after the side-scattered photon is recorded (second emission from the pair). Such an ``instability interval'' was absent in the succession between the two emissions focused on in the inset of Fig.~\ref{fig:FIG6}(b). The sign of $\langle A_{\rm \pi/2}\rangle_{\rm REC}$, resolved by the second emission from the pair after this short period of instability, is in contradiction to its expected value [since $h_{\rm \pi/2}(0)>0$ at the corresponding $\Delta\omega$]. The same pattern is observed from the bundle at $\kappa t \approx 39.57$: a spontaneous emission resolves the quantum beat after a very short ``instability interval'' conditioned on a cavity emission. None of there subtleties seem to concern us for the ME unraveling with $\theta=\pi/4$. When the LO is in phase with the direction of developing steady-state bimodality, only a single photon emission is recorded about the two-photon resonance peak, resolving a conditional field amplitude in agreement with the changing sign of $h_{\pi/4}(0)$ as the detuning is scanned [see Figs.~\ref{fig:FIG6}(b) and~\ref{fig:FIG8}(a)]. 
\newline

We conclude this section by reverting to the ensemble average evolution obeying the ME~\eqref{eq:ME}, but in the limit of ``zero system size''[$\gamma/(2\kappa)\to 0$], where we found that detailed balance is restored for the vacuum Rabi resonance. On this occasion we consider a coherent drive tuned to the peak of the two-photon resonance, and compute the intensity-field correlation function for the field quadrature $\theta=3\pi/4$. Figure~\ref{fig:FIG9} depicts $h_{3\pi/4}(\tau)$ for three increasing values of the drive amplitude resulting in a sub-photon steady-state cavity occupation. We observe that, as steady-state bimodality takes shape, the zero-delay intensity-field correlation changes sign, starting from a large-modulus negative value---similar to the behaviour of the squeezed quadrature amplitude in the OPO~\cite{GiantViolations2000}. Oscillations similar to those we encountered in Fig.~\ref{fig:FIG5}(b)[inset ii] occur at an intermediate timescale, associated with the imbalance between the two paths in the two-photon cascaded decay. Unlike the semiclassical oscillations of $h_{\theta}(\tau)$ at the vacuum Rabi resonance, the dominant oscillations are now independent of the drive amplitude, while superposed remnants of the quantum beat can be discerned. In parallel, the reduced variation in $h_{3\pi/4}(\tau)$ about unity is accompanied by a growing temporal asymmetry, signalling the breakdown of detailed balance. This result complements the temporal asymmetry noted in the cross-correlation between photons transmitted by the cavity when setting $r=1$ (direct photodetection at the APD), and those scattered out of its sides at the two-photon resonance peak~\cite{JOSAB2024}.

\section{Concluding remarks}

In summary, we have studied contextual correlations between the wave and particle aspects of light travelling down two different output channels, in a regime of cavity QED where quantum fluctuations produce a prevalent disagreement from mean-field predictions. Revealing layers of information inaccessible to the ME description, the wave-particle correlator unraveling details fluctuations which operationally resolve the JC spectrum, while exhibiting detuning-dependent nonclassical phase shifts in the selected quadrature amplitude of the electromagnetic field sustained by the resonator. The field amplitude is continuously monitored through conditional homodyne detection, making the part of the charge deposited on the detector circuit which is to be added to the shot noise component. On the other arm of the wave-particle correlator, the photon scattering rate in the ``start'' detector (APD) is brought closer to its ensemble average steady-state value when photoelectron ``click'' counts interrupt and reset the coherent evolution producing the photocurrent. As we approach the two-photon resonance peak, the manifestation of photon bunching depends on the setting of the local-oscillator phase, with the quantum beat playing the role of the system `clock'.  

The formal equivalence of `atomic' and field operators in the effective two-state model, employed here to calculate the intensity-field correlation via the quantum regression formula at the vacuum Rabi resonance, also places scattering into the cavity mode on an equal footing to scattering out of the sides of the cavity. The equivalence allows for mapping of the conditional and ensemble average dynamics to ordinary resonance fluorescence and the associated detailed balance imposed by the low dimensionality. Nevertheless, the numerical solution of the ME testifies to the departure from the two-state model, and to the excitation of dressed states higher up along the rungs of the JC ladder, which brings about the breakdown of detailed balance. When such states with asymmetric detunings are involved, the $a$ and $\sigma_{-}$ operators are no longer equivalent in terms of their dressed-state expansion. Hence, the developing ``spectral asymmetry'' in the photoemissive source is operationally translated into a temporal asymmetry of the extracted intensity-intensity and intensity-field correlations. Signs of temporal asymmetry are already notable when a drive on the order of magnitude of the dissipation rates is tuned to the vacuum Rabi resonance peaks. Time-asymmetric fluctuations become more ostensible at the two-photon resonance, in parallel with the onset of the dynamical instability accompanying a multilevel cascaded process. 

\bibliography{bibliographyFQMT}

\end{document}